\documentclass[prl,amssymb,amsmath,amsfonts,nofootinbib,reprint,showpacs,longbibliography]{revtex4-1}

\usepackage{graphicx}
\usepackage{lmodern}
\usepackage{amsmath,amssymb}
\usepackage{mathrsfs}
\usepackage{amsfonts}
\usepackage[utf8]{inputenc}
\usepackage{url}

\usepackage[dvipsnames]{xcolor}
\usepackage{multirow}
\usepackage[normalem]{ulem}
\usepackage{lipsum}

\usepackage{marvosym}
\usepackage{enumerate}
\usepackage{color,soul}
\usepackage{acronym}

\usepackage{bm}
\usepackage{color}
\usepackage{commath}
\allowdisplaybreaks
\usepackage{multirow}

\definecolor{linkcolor}{rgb}{0.1,0.4,0.8}
\usepackage[unicode, colorlinks=true, linkcolor=linkcolor, citecolor=linkcolor, filecolor=linkcolor, urlcolor=linkcolor, linktocpage, breaklinks]{hyperref}

\newcommand{\TEOB}[1]{\texttt{TEOBResumS{#1}}}
\newcommand{\NRPM}[1]{\texttt{NRPM{#1}}}

\newcommand{\orcid}[1]{\href{https://orcid.org/#1}{
\includegraphics[width=10pt]{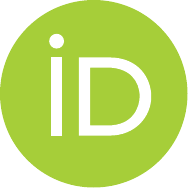}
}}

\def\CoRe{\texttt{CoRe}}

\def\BAM{\texttt{BAM}}

\def\Dali{\TEOB{-Dalí}}

\def\SEOBNRvE{\texttt{SEOBNRv5EHM}~}

\def\eg{\textit{e.g.}}
\def\ie{\textit{i.e.}}

\def\bbhsims{1183}
\def\bbhsimsprec{107} % precessing
\def\bnssims{27}
\def\bhnssims{185}

\newacro{adm}[ADM]{Arnowitt-Deser-Misner}
\newacro{bbh}[BBH]{binary black hole}
\newacro{bh}[BH]{black hole}
\newacroplural{bh}[BHs]{black holes}
\newacro{bhns}[BHNS]{black hole-neutron star}
\newacro{bns}[BNS]{binary neutron star}
\newacro{bf}[BF]{Bayes' factor}
\newacro{cbc}[CBC]{compact binary coalescence}
\newacro{ce}[CE]{Cosmic Explorer}
\newacro{da}[DA]{data analysis}
\newacro{et}[ET]{Einstein Telescope}
\newacro{emri}[EMRI]{extreme mass ratio inspiral}
\newacro{eob}[EOB]{effective-one-body}
\newacro{eos}[EoS]{equation of state}
\newacro{eom}[EOM]{equations of motion}
\newacro{fd}[FD]{frequency domain}
\newacro{fft}[FFT]{Fast Fourier transform}
\newacro{gw}[GW]{gravitational-wave}
\newacroplural{gw}[GWs]{Gravitational waves}
\newacro{gr}[GR]{general relativity}
\newacro{grb}[GRB]{gamma-ray burst}
\newacro{grhd}[GRHD]{general-relativistic hydrodynamics}
\newacro{gwosc}[GWOSC]{Gravitational Wave Open Science Center}
\newacro{gwtc1}[GWTC-1]{the first gravitational-wave transients catalog}
\newacro{gsf}[GSF]{Gravitational Self Force}
\newacro{hm}[HM]{Higher mode}
\newacroplural{hm}[HMs]{Higher modes}
\newacro{ifo}[IFO]{interferometer}
\newacro{imr}[IMR]{inspiral-merger-ringdown}
\newacro{im}[IMR]{inspiral-to-merger}
\newacro{kagra}[KAGRA]{Kamioka Gravitational Wave Detector}
\newacro{ligo}[LIGO]{Laser Interferometer Gravitational-Wave Observatory}
\newacro{lso}[LSO]{Last Stable Orbit}
\newacro{lvc}[LVC]{LIGO-Virgo Collaboration}
\newacro{lvk}[LVK]{LIGO-Virgo-Kagra Collaboration}
\newacro{lo}[LO]{leading order}
\newacro{ns}[NS]{neutron star}
\newacroplural{ns}[NSs]{neutron stars}
\newacro{nr}[NR]{numerical relativity}
\newacro{nqc}[NQCs]{next-to-quasicircular corrections}
\newacro{nlo}[NLO]{next-to-leading order}
\newacro{nnlo}[NNLO]{next-to-next-to-leading order}
\newacro{n3lo}[N3LO]{next-to-next-to-next-to-leading order}
\newacro{n4lo}[N3LO]{next-to-next-to-next-to-next-to-leading order}
\newacro{ode}[ODE]{Ordinary Differential Equation}
\newacroplural{ode}[ODEs]{Ordinary Differential Equations}
\newacro{pn}[PN]{post-Newtonian}
\newacro{pm}[PM]{post-Minkowskian}
\newacro{pe}[PE]{parameter estimation}
\newacro{psd}[PSD]{power spectral density}
\newacro{pa}[PA]{post-adiabatic}
\newacro{qnm}[QNM]{quasi-normal mode}
\newacro{qc}[QC]{quasi-circular}
\newacro{rwz}[RWZ]{Regge-Wheeler-Zerilli}
\newacro{sm}[SM]{Supplemental Material}
\newacro{snr}[SNR]{signal-to-noise ratio}
\newacro{spa}[SPA]{stationary-phase approximation}
\newacro{sxs}[SXS]{Simulating eXtreme Spacetimes}
\newacro{td}[TD]{time domain}
\newacro{ng}[NG]{Next Generation}

\definecolor{cyan}{rgb}{0,0.9,0.9}
\definecolor{orange}{rgb}{0.9,0.5,0}
\definecolor{magenta}{rgb}{1,0,1}
\definecolor{purple}{rgb}{0.8,0.4,0.8}
\definecolor{gray}{rgb}{0.8242,0.8242,0.8242}
\definecolor{dodgerblue}{rgb}{0.12, 0.56, 1.0}
\definecolor{forestgreen}{rgb}{0.1, 0.8, 0.3}

\begin{document}

\title{Effective-one-body modeling for generic compact binaries with arbitrary orbits}

\author{Simone \surname{Albanesi}\orcid{0000-0001-7345-4415}$^{1,2}$}
\author{Rossella \surname{Gamba}\orcid{0000-0001-7239-0659}$^{3,4}$}
\author{Sebastiano \surname{Bernuzzi}\orcid{0000-0002-2334-0935}$^{1}$}
\author{Joan \surname{Fontbuté}\orcid{0009-0004-7893-7386}$^{1,5}$}
\author{Alejandra \surname{Gonzalez}\orcid{0000-0002-5034-9353}$^{1,6}$}
\author{Alessandro \surname{Nagar}\orcid{0000-0001-7998-2673}$^{2,7}$}
\affiliation{${}^{1}$ Theoretisch-Physikalisches Institut, Friedrich-Schiller-Universit{\"a}t Jena, 07743, Jena, Germany}
\affiliation{${}^{2}$ INFN Sezione di Torino, Via P. Giuria 1, 10125 Torino, Italy}
\affiliation{${}^{3}$ Institute for Gravitation \& the Cosmos, The Pennsylvania State University, University Park PA 16802, USA}
\affiliation{${}^{4}$ Department of Physics, University of California, Berkeley, CA 94720, USA}
\affiliation{${}^{5}$ Departament de F{\'\i}sica Qu\`antica i Astrof\'{\i}sica, Institut de Ci\`encies del Cosmos, Universitat de Barcelona, Mart\'{\i} i Franqu\`es 1, E-08028 Barcelona, Spain}
\affiliation{${}^{6}$ Institut d’Aplicacions Computacionals i de Codi Comunitari (IAC3) - IEEC, Universitat de les Illes Balears, Crta. Valldemossa km 7.5, E-07122 Palma, Spain}
\affiliation{${}^{7}$Institut des Hautes Etudes Scientifiques, 91440 Bures-sur-Yvette, France}

\begin{abstract}
    We present the first unified model for the general relativistic dynamics
    and gravitational radiation of generic compact binaries. 
    \Dali{} is a model based on the effective-one-body framework incorporating
    tidal interactions, generic spins, multipolar radiation reaction/waveform and
    numerical-relativity information. 
    It allows the computation of gravitational waves and other dynamical gauge
    invariants from 
    \textit{generic} binaries (black holes, neutron stars, neutron
    star-black hole binaries) evolving along \textit{arbitrary}
    orbits (quasi-circular, eccentric, non-planar) through merger and
    including scattering.
    The performances of \Dali{} in the strong-field
    regime are showcased by comparisons with a large sample of 1395 high-accuracy
    numerical-relativity simulations available.  
    \Dali{} allows the computation of faithful waveforms for gravitational wave astronomy, 
    providing at the same time an understanding and a prediction of the strong-field dynamics.
    
\end{abstract}

\date{\today}
\maketitle

% reset all acronyms
\acresetall

\noindent {\textbf{\textit{Introduction.}}}
\acp{gw} from compact binaries have revolutionized our understanding of the Universe over the past few years~\cite{LIGOScientific:2018mvr,LIGOScientific:2020ibl,LIGOScientific:2021djp}.
The detection and parameter estimation for current and future \ac{gw} observatories~\cite{LIGOScientific:2014pky,VIRGO:2014yos,Punturo:2010zz,Maggiore:2019uih,Reitze:2019iox,LISA:2017pwj} 
require models able to generate waveforms for compact binaries. 
Central to this endeavour is the accurate solution of the relativistic two-body problem
for different kinds of binaries and arbitrary orbits.  
The \ac{eob} formalism provides us with a powerful tool to describe waveforms and the underlying 
dynamics~\cite{Buonanno:1998gg,Buonanno:2000ef,Damour:2000we}. 
Its key elements are a Hamiltonian
and a radiation reaction force, which is responsible for the energy and 
angular momentum losses. Both these elements incorporate
\ac{pn}, \ac{nr} and \ac{gsf} (and test-mass) information in a resummed way, to make the 
model predictive in the high-frequency regime. 
While initially developed for quasi-circular non-spinning
\acp{bbh}~\cite{Damour:2007xr,Buonanno:2006ui,Damour:2007vq,Buonanno:2007pf,Damour:2008gu,Buonanno:2009qa,Damour:2009kr}, 
the \ac{eob} approach has been extended to include spin interactions~\cite{Damour:2001tu,Pan:2009wj,Barausse:2009xi,Pan:2010hz,Barausse:2011ys,Taracchini:2012ig,Pan:2013rra,Taracchini:2013rva,Damour:2014sva}, 
subdominant modes~\cite{Pan:2011gk,Cotesta:2018fcv,Nagar:2019wds,Nagar:2020pcj,Riemenschneider:2021ppj,Nagar:2023zxh,Pompili:2023tna},
tidal interactions~\cite{Damour:2009vw,Damour:2009wj,Damour:2012yf,Bernuzzi:2014kca,Hinderer:2016eia,Steinhoff:2016rfi,Nagar:2018zoe,Lackey:2018zvw,Gamba:2020ljo,Steinhoff:2021dsn,Gonzalez:2022prs,Gamba:2022mgx,Gamba:2023mww}, 
precessing systems~\cite{Buonanno:2005xu,Pan:2013rra,Akcay:2020qrj,Ossokine:2020kjp,Gamba:2021ydi,Khalil:2023kep,Ramos-Buades:2023ehm,Gamba:2024cvy}, and
arbitrary planar orbits~\cite{Hinderer:2017jcs,Chiaramello:2020ehz,Nagar:2020xsk,Khalil:2021txt,Placidi:2021rkh,Ramos-Buades:2021adz,Albanesi:2022xge,Placidi:2023ofj,Albanesi:2023bgi,Nagar:2024oyk,Gamba:2021gap,Andrade:2023trh,Albanesi:2024xus,Grilli:2024lfh,Gamboa:2024hli,Gamboa:2024hli}.
The EOB description of the inspiral-plunge phase is then completed with post-merger models
for the vacuum~\cite{Damour:2014yha} and matter~\cite{Breschi:2022xnc,Gonzalez:2022prs} cases, 
thus providing a description of the whole evolution.
\begin{figure*}[th]
  \centering 
    \includegraphics[width=\textwidth]{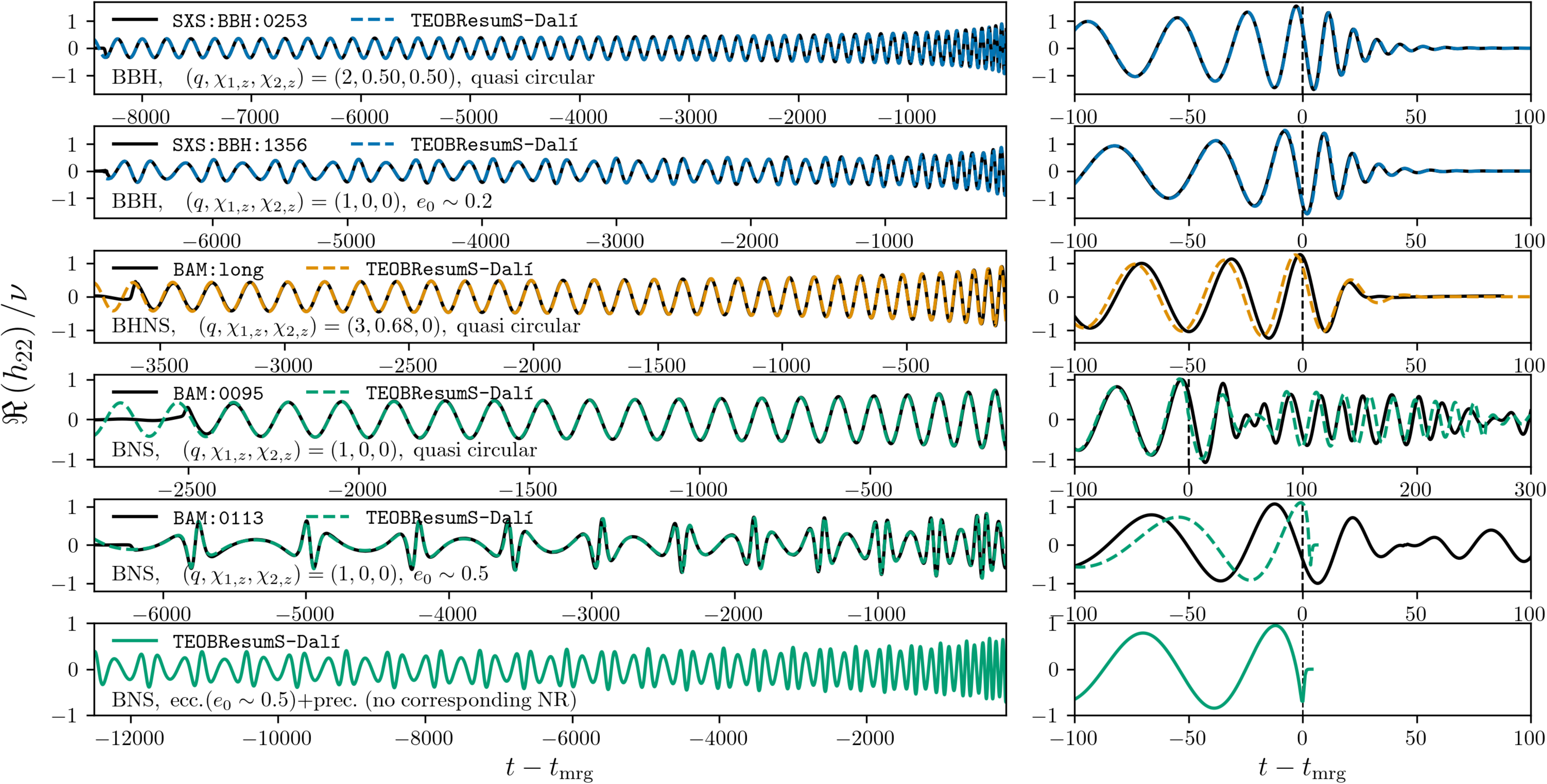}
    \caption{\ac{eob} and \ac{nr} waveforms for a sample of coalescing binaries. Our comparisons 
    cover quasi-circular and eccentric spin-aligned BBHs (blue) and BNSs (green) and quasi-circular BHNS systems (orange). 
    Waveforms are aligned during the inspiral. For the eccentric and precessing BNS system, 
    we present only the \ac{eob} waveform (green), as no corresponding \ac{nr} data is currently available. 
    }
  \label{fig:waves}
\end{figure*}
Despite significant progress, no EOB model incorporate all these effects throughout the entire coalescence as presented in this work.
The availability of a unified EOB model 
has important consequences on our capabilities to learn additional information from \ac{gw} data: 
the incomplete description of \ac{gw} sources can lead to systematic errors in the estimation of their parameters,
limiting the scientific return of observations~\cite{Divyajyoti:2023rht}.

This was demonstrated in the analyses of the GW190521 event~\cite{LIGOScientific:2020iuh} conducted in
Refs.~\cite{CalderonBustillo:2020fyi, Gayathri:2020coq, Gamba:2021gap}, which respectively favored a head on, higly eccentric or dynamical
capture scenario over a quasi-spherical one, 
highlighting the importance of models that accurately describe arbitrary orbits to avoid missing key physical insights.
Similarly, Bayesian analyses of GW190814~\cite{LIGOScientific:2020zkf} using the previous \TEOB{} unified model 
for \ac{bbh} and \ac{bhns} has reinforced the decisive evidence for the presence of higher modes~\cite{Gonzalez:2022prs}. 
Such model can be crucial to infer the presence of a neutron star in the source in future events.

Here, we discuss \Dali{}, the first EOB model 
offering a comprehensive description of generic compact binary dynamics, along with the corresponding 
waveforms.
\Dali{} allows the precise computation of \acp{gw} from \acp{bbh}, \acp{bns} and \ac{bhns} evolving along any orbit (quasi-circular, eccentric, hyperbolic, non-planar),
from arbitrarily large separations (low frequencies) to merger. 
The waveform is completed with ringdown for \ac{bbh}~\cite{Damour:2014yha,Nagar:2020pcj} and
\ac{bhns}~\cite{Zappa:2019ntl,Gonzalez:2022prs}, and post-merger emission for \ac{bns}~\cite{Breschi:2022xnc}.
%the latter has been also used to analyse GW170817~\cite{LIGOScientific:2017vwq}.  
We discuss the model's performance by comparing with a sample of high-accuracy \ac{nr} 
simulations: \bbhsims{} \acp{bbh}, \bnssims{} \acp{bns}, and \bhnssims{} \ac{bhns} binaries.
This comparison represents one of the most exhaustive tests of any \ac{gw} model to date, in terms of
the variety of configurations considered and the number of \ac{nr} simulations used.
We focus on faithfulness analysis of the \ac{gw} signal to showcase, with a global measure, the 
accuracy of the model, and we consider scattering angles for open orbits. We further showcase 
how the \ac{eob} can be employed to inform future numerical studies of systems with \acp{ns}.

\noindent {\textbf{\textit{Conventions.}}} We denote the total rest mass of the binary as $M = m_1 + m_2$, 
the symmetric mass ratio as $\nu = m_1 m_2 / M^2$, the mass ratio as $q = m_1/m_2 \geq 1$,
the dimensionless spins as $\chi_{1,2} = S_{1,2}/(m_{1,2}^2)$ and the dimensionless 
precession spin parameter as $\chi_p = \max \left( |\bm{\chi_{1,\perp}}|, \frac{4+3q}{4q^2+3q} |\bm{\chi_{2,\perp}}| \right)$~\cite{Schmidt:2014iyl}.
We also use $p_\varphi \equiv L/(\nu M^2)$, 
where $L$ is the orbital angular momentum. 
The quadrupolar dimensionless tidal polarizability parameters for
gravitoelectric interactions are indicated as $\Lambda_{1,2}$.
We denote the reduced tidal parameter as $\tilde\Lambda=16/13\,
(m_1+12m_2)m_1^4\Lambda_1/M +
(1\leftrightarrow2)$~\cite{Favata:2013rwa}, which is equivalent to
the tidal coupling constant introduced in~\cite{Damour:2009wj} for
equal-mass binaries. 
The \ac{gw} polarizations $h_{+,\times}$ are decomposed as
$ h_+ - i h_\times = {\cal D}_L^{-1}\sum_{\ell=2}^{\infty}
\sum_{m=-\ell}^{\ell} h_{\ell m} {}_{-2}Y_{\ell m}(\iota,\varphi)$,
where ${\cal D}_L$ is the luminosity distance, $\iota$ the inclination
angle, $\varphi$ the orbital reference phase,  
${}_{-2}Y_{\ell m}$ are the spin-weighted spherical harmonics
and $h_{\ell m} = A_{\ell m} e^{-i \phi_{\ell m}}$ are the \ac{gw} modes.
%Unless otherwise stated, 
We use geometric units with $G = c = 1$.

\noindent {\textbf{\textit{\Dali{}.}}}

To model \acp{bbh}, \Dali{} uses a 5PN Hamiltonian with two parameters 
calibrated to \ac{nr} quasi-circular simulations and a $3^{+3}$\ac{pn} description of all multipoles\footnote{This notation means that
the waveform multipoles contain $\nu$-dependent information up to $3$\ac{pn} order, hybridized with test-mass terms as to reach (formal) 
$6$\ac{pn} accuracy.} up to $\ell=8$ with the exception of the $(2,2)$, taken at full $4$\ac{pn} order~\cite{Nagar:2024oyk}. 
Spin-spin couplings are incorporated using the concept of centrifugal radius~\cite{Damour:2014sva}, $r_c$, 
including up to \ac{nlo} PN information.
Predictability in the strong-field regime is obtained by the systematic use of Pad\'e resummations~\cite{Messina:2018ghh}.
In particular, the log-terms of the residual amplitudes in the resummed radiation of Ref.~\cite{Damour:2008gu} are treated as 
proposed in Ref.~\cite{Nagar:2024oyk}. 
Analytical non-circular corrections to the azimuthal radiation reaction force are accounted for via a 
arbitrary-orbits Newtonian prefactor, which explicitly depends on first and second time derivatives of the orbital 
frequency and up to third time derivatives of the radial separation~\cite{Chiaramello:2020ehz,Nagar:2024oyk}.
Energy losses due to radial dissipation, instead, are included by factorizing the angular momentum flux~\cite{Nagar:2023zxh}.
The validity of this prescription has also been confirmed in the test-mass limit~\cite{Albanesi:2021rby,Albanesi:2022ywx}, 
as this regime provides a well-controlled setting for testing analytical
prescriptions for comparable-mass \ac{eob} models~\cite{Nagar:2006xv,Yunes:2009ef,Yunes:2010zj,Barausse:2011kb,Bernuzzi:2010ty,Bernuzzi:2010xj,Bernuzzi:2011aj,Bernuzzi:2012ku,Taracchini:2013wfa,Taracchini:2014zpa,Albanesi:2021rby,Albanesi:2022ywx,Faggioli:2024ugn}.
Effects due to spin-induced precession are accounted for via a hybrid \ac{pn}-\ac{eob} scheme~\cite{Akcay:2020qrj,Gamba:2021ydi,Gamba:2024cvy},
which relies on the timescale separation between the precession and the orbital motion~\cite{Schmidt:2010it,Boyle:2011gg,Schmidt:2012rh,Hannam:2013oca,Schmidt:2014iyl}.
The inspiral-plunge waveform is completed with 
a quasi-circular \ac{nr}-informed ringdown model~\cite{Damour:2014yha,Nagar:2021gss}.
The smooth transition is aided using \ac{nr}-informed \ac{nqc}~\cite{Damour:2007xr,Riemenschneider:2021ppj} 
in the quasi-circular and mildly eccentric cases.
Spin-dependent horizon absorption effects are included to \ac{lo}~\cite{Damour:2014sva}.

\ac{bns} are described using similar analytical prescriptions to \ac{bbh} but adding tidal interactions. The latter are incorporated in the Hamiltonian by including all the gravitoelectric and gravitomagnetic terms currently known and up to multipolar order $\ell=8$.
These interactions are parametrized by the usual multipolar set of tidal polarizability parameters; the dominant (gravitoelectric, $\ell=2$) contribution is parametrized by the reduced tidal parameter $\tilde\Lambda$~\cite{Damour:2009wj}.
Different prescriptions for the tides are available: tidal potentials in \ac{pn} form \cite{Bernuzzi:2012ci}, \ac{gsf}-resummed expressions of the leading order multipolar terms~\cite{Bini:2014zxa,Bernuzzi:2014owa,Akcay:2018yyh} and a $f$-mode resonance model~\cite{Gamba:2022mgx}. EOB/NR comparisons with precision NR data suggest that the \ac{gsf}-resummed potential performs best \cite{Bernuzzi:2014owa,Akcay:2018yyh,Gamba:2022mgx}. Hence, the latter remains the default choice also in \Dali{}.
Nonlinear-in-spin interactions dependent on finite-size effects are
incoporated up to \ac{nnlo} in a PN-resummed fashion using the
centrifugal radius and an effective spin variable
\cite{Nagar:2018zoe}.
The multipolar waveform incorporates known \ac{pn} coefficients up to $2$\ac{pn} order~\cite{Gamba:2022mgx}.
A postmerger completion that accounts for the emission from the NS remnant is obtained using NR-informed analytical 
models \NRPM{}~\cite{Breschi:2019srl} or \NRPM{w}~\cite{Breschi:2022xnc,Breschi:2022ens}. 
These models leverage on a quasi-universal character of \ac{bns} postmerger GWs that allows a parametrization of the wave in terms of the same tidal polarizability parameters employed in the EOB Hamiltonian~\cite{Bernuzzi:2015rla}. Therefore, \Dali{} delivers a robust representation of the complete BNS spectrum~\cite{Breschi:2019srl}, which can be used for inference on extreme matter effects with next-generation observations, \eg~\cite{Breschi:2021xrx,Prakash:2023afe,Fields:2023bhs}.

\ac{bhns} are described by incorporating the same tidal interactions as for \ac{bns} in the 
Hamiltonian but adopting \ac{nqc} and ringdown model similar to \ac{bbh}. 
The approach is the same as the one described in Ref.~\cite{Gonzalez:2022prs}, but the \Dali{} 
implementation improves the multipolar \ac{nqc} and the remnant+ringdown model by 
the use of new numerical data from the \CoRe~collaboration~\cite{Gonzalezinprep}. 
The remnant \ac{bh} from \ac{bhns} can be different from those of \ac{bbh} because of the tidal disruption of the \ac{ns} 
during merger. Therefore, the remnant's mass and spin are modeled from \ac{nr} data using an analytical representation 
that smoothly deforms the BBH case \cite{Zappa:2019ntl}. Similarly, \ac{nqc} to the waveform 
for higher modes up to $\ell=4$ have been developed from the new simulations together with a smooth deformation of the 
ringdown model used for \ac{bbh}. 

On the algorithmic point of view, \Dali{} inherited from previous \TEOB{} models two key analytical 
acceleration techniques (for quasi-circular dynamics): the post-adiabatic approximation for the fast 
solution of the Hamiltonian equations of motions~\cite{Nagar:2018gnk} and the EOBSPA for the computation 
of frequency-domain EOB waveforms~\cite{Gamba:2020ljo}.

We showcase  time-domain waveform comparisons with \ac{nr} in Fig.~\ref{fig:waves}.
From top to bottom we consider:
a quasi-circular non-spinning \ac{bbh} $q=2$ case from the SXS catalog~\cite{SXS:catalog} , 
a mildly eccentric \ac{bbh},
a quasi-circular \ac{bhns} case~\cite{Gonzalezinprep}, 
a quasi-circular and a highly eccentric \acp{bns} from the \CoRe{} catalog~\cite{Dietrich:2018phi,Gonzalez:2022mgo}.
The last panel shows an \ac{eob} waveform for an
eccentric and precessing \ac{bns} ($e_0^{\rm EOB} = 0.5$, $\chi_{1}=(0.1, 0.3, 0.1)$,  $\chi_{2}=(0.1, 0.0, 0.4)$), 
which have not been explored so far in \ac{nr}.
The EOB/NR agreement is striking for all the configurations 
considered. The consistency of both phase and amplitude is maintained over the 
duration of the entire \ac{nr} signals. 
Indeed, the quasi-circular spin-aligned \ac{bbh} in the top row of Fig.~\ref{fig:waves}
yields an excellent phase difference at merger ($\Delta \phi^{\rm mrg}_{22} \simeq 0.057$~rad).
For the eccentric \ac{bbh} we have, instead,
$\Delta \phi^{\rm mrg}_{22} \simeq 0.073$~rad,
a value that is only slighty higher than the one obtained for 
the standard equal mass configuration \texttt{SXS:BBH:0180} 
($\Delta \phi^{\rm mrg}_{22} \simeq 0.036$~rad, not shown in Fig.~\ref{fig:waves}).
For the quasi-circular \ac{bhns} case, we get a dephasing of $\Delta \phi^{\rm mrg}_{22} \simeq 0.502$~rad, a value
similar to the quasi-circular \ac{bns} here considered ($\Delta \phi^{\rm mrg}_{22} \simeq 0.579$~rad).
However, these dephasings are compatible with \ac{nr} uncertainties. 

\noindent {\textbf{\textit{Systematic EOB/NR comparisons.}}}
\begin{figure}[t]
  \centering 
    \includegraphics[width=0.49\textwidth]{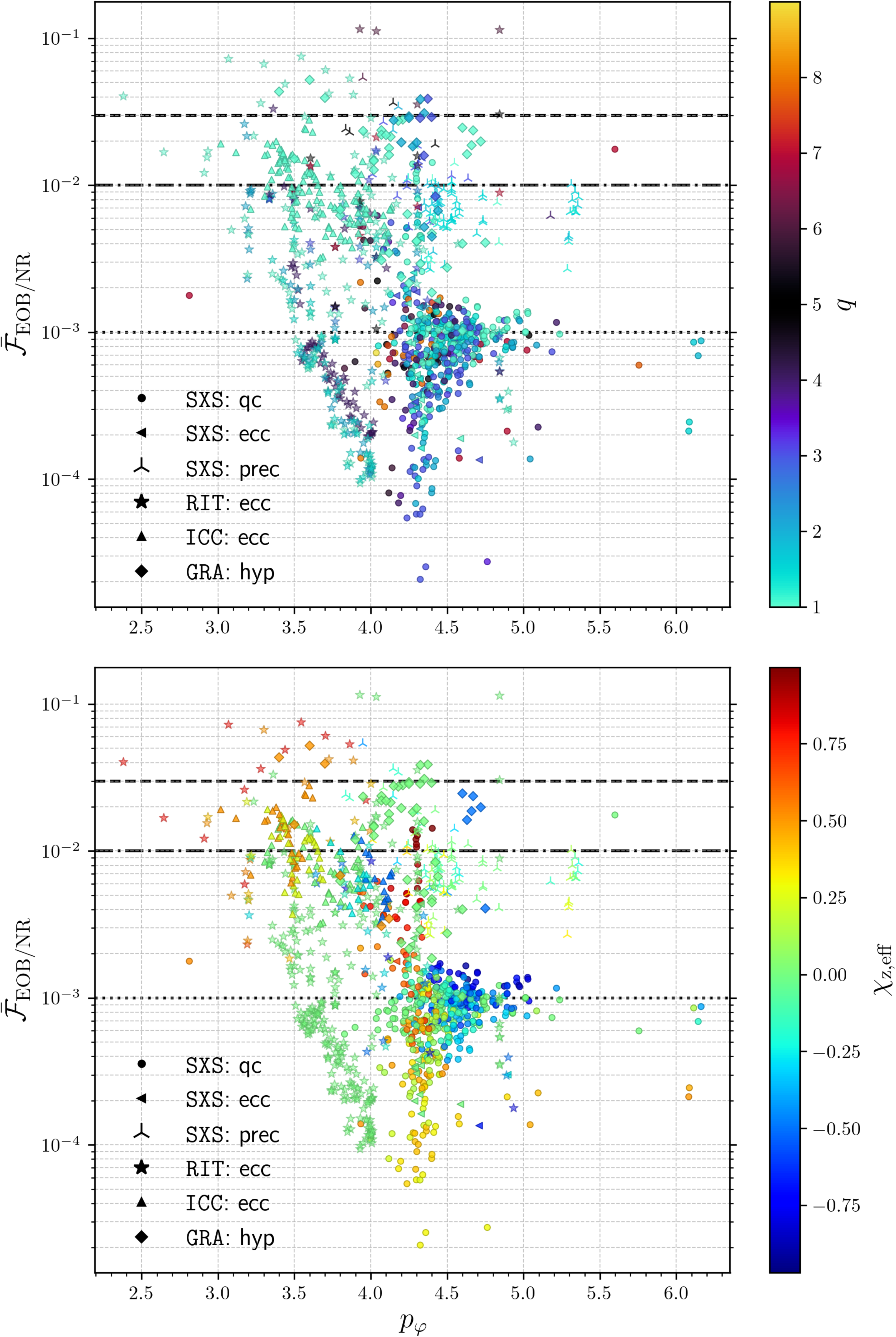}
    \caption{
    EOB/NR mismatches for the \bbhsims{}  \acp{bbh} considered in this work.
    The colormaps highlight the mass ratio(upper panel) and the (initial) $z$-component of the effective spin (lower panel). 
    Horizontal lines mark the $0.1\%$, $1\%$ and $3\%$ thresholds. }  
    \label{fig:mm_bbh}
\end{figure}
We consider a faithfulness analysis with \ac{nr} data for a systematic accuracy assessment.
The \ac{eob}/\ac{nr} mismatch (or unfaithfulness) $\bar{\mathcal{F}}$ is defined as 
the noise-weighted inner product of the two frequency-domain waveforms, $h_{\rm EOB}$ and
$h_{\rm NR}$:
\begin{equation}
  \bar{\mathcal{F}} = 1 - \max_{t_{\rm ref}, \phi_{\rm ref}, \theta_a} \mathcal{O}(h_{\rm NR}, h_{\rm EOB})\, ,
\end{equation}
where $\mathcal{O}$ is the overlap integral\footnote{If subdominant multipoles
are employed, the sky-maximized overlap statistic should be considered, see \eg~Ref.~\cite{Harry:2017weg}.}
with the detector's \ac{psd} $S_n(f)$
\begin{equation}
  \mathcal{O}(h_{\rm NR}, h_{\rm EOB}) = 4 \Re\int_{f_0}^{f_{max}} \frac{\tilde{h}_{\rm NR} \tilde{h}_{\rm EOB}^*}{S_n(f)} df\, .
\end{equation}
We use the zero-detuned, high-power noise spectral density of Advanced LIGO~\cite{aLIGODesign_PSD}.
The maximization is performed over a reference time $t_{\rm ref}$ and phase $\phi_{\rm ref}$,
and an additional set of parameters $\theta_a$ that depend on the specific nature of the configuration considered.
For instance, following Ref.~\cite{Pratten:2020ceb}, for quasi-spherical (spin-precessing) systems $\theta_a = \{\varphi_{\rm plane}\}$, 
where $\varphi_{\rm plane}$ is an angle that rotates the in-plane components of the spins at a reference frequency.
For eccentric or hyperbolic systems, $\theta_a = \{E^0, p_\varphi^0\}$ are the initial energy and angular momentum, respectively.
Then initial separation is set as a large radius in the unbound case, while it is set to the apastron 
identified by $(E_0, p_\varphi^0)$ for bound cases. 
To compute the mismatches, we use the package \texttt{pycbc}~\cite{Biwer:2018osg}.
In the case of scattering orbits, \ac{nr} waveforms are typically not yet sufficiently accurate; 
we have thus also performed an analysis of the scattering angle $\theta$ following Ref.~\cite{Damour:2014afa} and subsequent works~\cite{Damour:2022ybd,Rettegno:2023ghr,Albanesi:2024xus,Fontbute:2024amb,Swain:2024ngs}.

\noindent {\textbf{\textit{Binary black holes.}}}
We consider \ac{nr} waveforms from several codes, totaling \bbhsims{} configurations.
Out of these, 635 are quasi-circular (528) or quasi-spherical (\bbhsimsprec) simulations from the SXS 
catalog~\cite{Chu:2009md,Lovelace:2010ne,Lovelace:2011nu,Buchman:2012dw,Hemberger:2013hsa,Scheel:2014ina,Blackman:2015pia,Lovelace:2014twa,Mroue:2013xna,Kumar:2015tha,Chu:2015kft,Boyle:2019kee,SXS:catalog}; 
28 are the eccentric public simulations from the same SXS catalog;
338 are eccentric binaries from the RIT catalog~\cite{Healy:2020vre,Healy:2022wdn},
128 correspond to highly eccentric configurations of the \texttt{ICC} catalog~\cite{Andrade:2023trh,data1297_2025} 
and, finally, 54 are dynamical captures and scatterings~\cite{Albanesi:2024xus} 
from the \CoRe{} collaboration performed with \texttt{GR-Athena++}~\cite{Daszuta:2021ecf,Rashti:2023wfe,Daszuta:2024ucu}.

Mismatches are computed with the dominant $(2,|2|)$ mode for spin-aligned \acp{bbh} and with the
$(2,|2|), (2,|1|), (3,|3|)$ and $(4,|4|)$ co-precessing modes for precessing systems. 
For the latter, we consider an inclination of $\pi/3$ and average over the phase and effective polarization 
angle. 
They are weighted with the zero-detuned, high-power noise spectral density of Advanced
LIGO~\cite{aLIGODesign_PSD}.
The mismatches that we show have been computed for a reference mass of $100 M_\odot$ in the frequency band $\left[10,1024\right] \rm Hz$.
The RIT catalog contains several almost head-on configurations, and therefore we consider only systems with 
$p_\varphi^0\geq p_\varphi^{\rm LSO}$, where $p_\varphi^{\rm LSO}$ is the angular momentum that corresponds to the \ac{lso}, which depends 
on the spins and mass ratio of the \ac{bbh}.
Following the discussion in Ref.~\cite{Albanesi:2024xus}, for dynamical captures 
we consider the configurations with $E_0\lesssim 1.2\,M$.

% Bound BBHs
Our results are reported in Fig.~\ref{fig:mm_bbh}. The two panels show the same mismatches plotted against the 
orbital angular momentum of the system, while the two colormaps indicate the mass ratio and 
the effective spin\footnote{For the precessing cases, the color highlights the initial $z$-component.}.
The $86.3\%$ of the configurations lead to a mismatch below $0.01$, 
while $97.9\%$ ($99.2\%$) of them are below the $0.03$ ($0.05$) threshold. 
The largest mismatches are obtained for configuration with high mass-ratio, low angular momentum and/or large in-plane spins.
This is expected, since in these regions the \ac{nr} data are more sparse and \Dali{} is not \ac{nr}-informed there.
Nonetheless, \Dali{} accurately reproduces the \ac{nr} waveforms 
for the vast majority of the configurations considered.
Note that mismatches of $10^{-4}$ graze the numerical uncertainties of the data. 

The recent model \SEOBNRvE{}~\cite{Gamboa:2024hli,Gamboa:2024imd}, 
which adopts non-circular corrections restricted to the bound case rather than
for arbitrary planar orbits, 
shows some improved mismatches with a set of elliptic SXS simulation (see, \eg, Fig.~4 and~13 of \cite{Gamboa:2024hli}).
While we cannot verify this with our latest model, since these simulations are not publicly available, 
we stress that \Dali{} is application ready for arbitrary motion (hyperbolic, elliptic, non-planar) and matter effects. 

% scatterings
Previous works showed that \Dali{} can be employed to quantitatively predict
general relativistic dynamics for dynamical encounters (initially open orbits that
become bound due to radiation reaction) and scatterings~\cite{Nagar:2020xsk}, including the transition 
between the two regimes~\cite{Albanesi:2024xus}. 
Therefore, we also compare \ac{nr} scattering orbits 
from an extended set of simulations, considering 42 additional configurations from Ref.~\cite{Rettegno:2023ghr} and 69 from Ref.~\cite{Fontbute:2024amb}, finding that \Dali{} reproduces accurately \ac{nr} results up to a few percent 
for low scattering angles (see \ac{sm}).

\noindent {\textbf{\textit{Binaries with neutron stars.}}}
\begin{figure}[t]
  \centering 
    \includegraphics[width=0.49\textwidth]{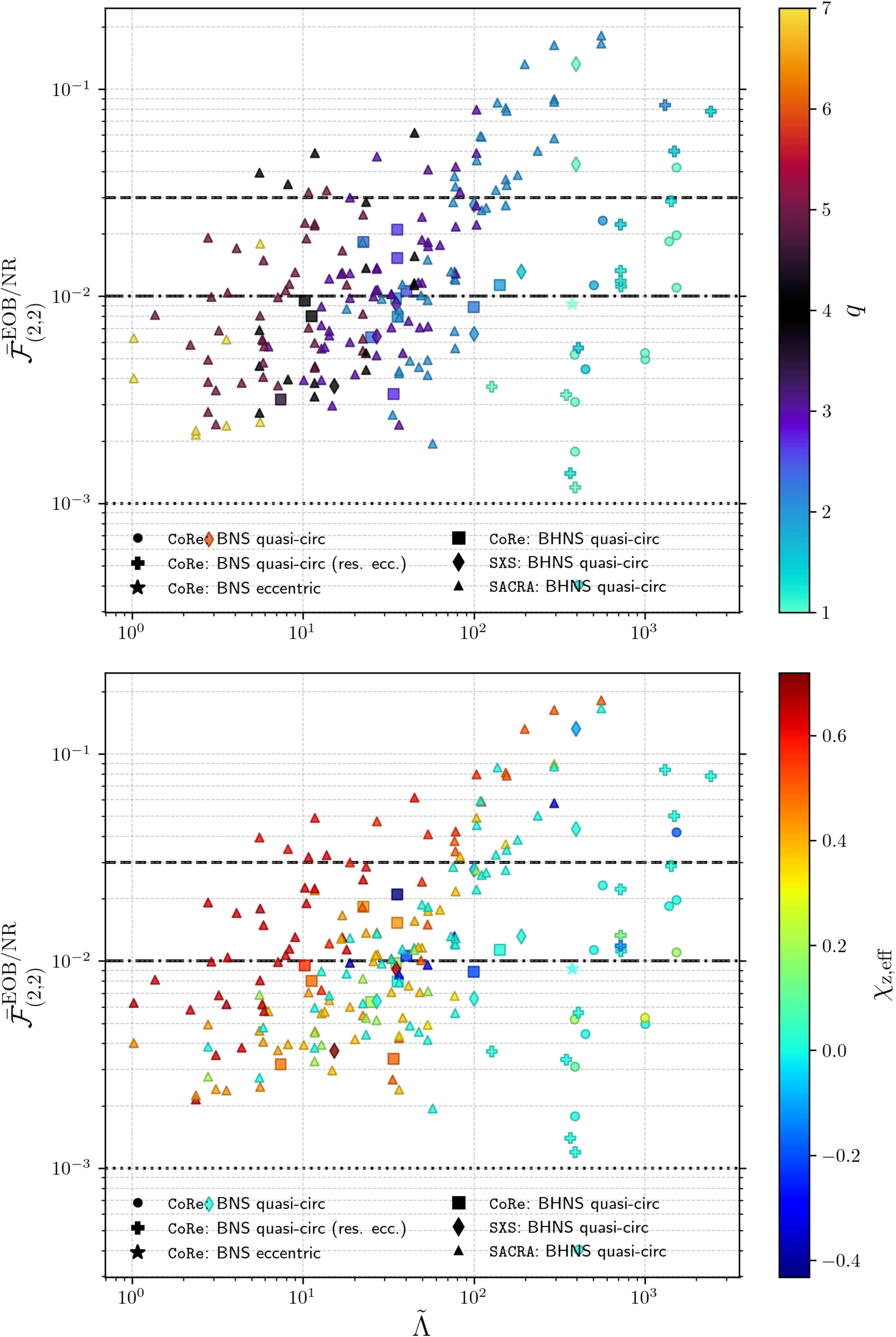}
    \caption{EOB/NR mismatches computed with the (2,2) mode for \ac{bns} and
    \ac{bhns} data from different catalogs. The lines mark the $0.1\%$, $1\%$ and $3\%$ thresholds.
    }
  \label{fig:mm_matter}
\end{figure}
For \ac{bns} we consider 26 quasi-circular (13 of which with residual eccentricity ${\lesssim} 10^{-4}$) and one eccentric \ac{bns} from the \CoRe{} database. 
The chosen data are representative of the longest and most accurate simulations available to date.
We report the corresponding EOB/NR mismatches for the (2,2) mode up to merger in Fig.~\ref{fig:mm_matter}.
Note that higher mismatches are found for high values of $\tilde{\Lambda}$,
which are incompatible with GW170817~\cite{LIGOScientific:2017vwq}. 
The eccentric configuration considered
corresponds to an initial EOB eccentricity $e_0^{\rm EOB} \sim 0.58$ (see Fig.~\ref{fig:waves}) and 
yields a mismatch of $0.9\%$. This highlights the accuracy of \Dali{} for eccentric \ac{bns}~\cite{Gamba:2022mgx}. 
\ac{bns} waveforms generated with \Dali{} can also be completed with the post-merger model
\texttt{NRPMw}, which has been tested against 102 independent \ac{nr} simulations not used to design the model~\cite{Breschi:2022xnc}.
The precision favorably compares against other post-merger models~\cite{Tsang:2019esi},
and is compatible with \ac{nr} uncertainties.

% BHNSs
For \ac{bhns} systems, we consider 162 spin-aligned simulations from the SACRA catalog~\cite{Kyutoku:2010zd,Kyutoku:2015gda}, 
9 from the SXS collaboration~\cite{duez_matthew_2019_3311687,
kidder_larry_2019_3311707,
sxs_collaboration_2019_3311678,
kidder_larry_2019_3337271,
foucart_francois_2019_3337268,
buonanno_alessandra_2019_3337269,
hinderer_tanja_2019_3337270,
foucart_francois_2020_4139881,
duez_matthew_2020_4139890}, 
and 13 new simulations from the \CoRe{} collaboration~\cite{Gonzalezinprep}. 
The mismatches for the inspiral-merger-ringdown (2,2) mode in the frequency band $\left[10,4096\right] \rm Hz$
computed using \Dali{} are reported in Fig.~\ref{fig:mm_matter}. 
Similarly to the \ac{bns} case, higher mismatches are obtained for configurations with 
large $\tilde{\Lambda}$.
Regarding the SXS data, the two highest mismatches are obtained for 
\texttt{SXS:BHNS:0005} ($13.2\%$) and \texttt{SXS:BHNS:0007} ($2.8\%$), which have spinning \acp{ns}. 
Some high mismatches are also obtained for some SACRA configurations, but it should be noted that
these are short simulations and the \ac{nr} uncertainty is large. 
Regarding the new \BAM{} \ac{bhns} simulations~\cite{Gonzalezinprep}, 
we consider the ones which have at least $500\,M$ of evolution before the merger.
The longest \BAM{} simulation available ($3500\, M$ before merger), which has $q=2.7$, a \ac{bh} with $\chi_z=0.68$ and non-spinning 
\ac{ns} with $\Lambda=599$, yields a mismatch of $\sim 0.3\%$. 
Overall, the model remains accurate for \ac{bhns} binaries with astrophysically relevant tidal deformabilities. 
We finally consider an additional precessing $q=2.5$ configuration computed with \BAM{}, 
which has a precessing spin $\chi_p=0.61$, a \ac{ns} with $\Lambda\simeq 494$, and yields a mismatch of $1.4\%$.

% Greedy    
Binaries with \acp{ns} are less investigated in \ac{nr} due to the higher computational cost and the increased dimensionality of its 
parameter space compared to \ac{bbh} systems. \Dali{} allows us to identify which regions of the parameter 
space for these systems should be prioritized for future \ac{nr} simulations. 
We perform a greedy search based on a parameter space defined by masses, spins and $\Lambda_{1,2}$
and using \Dali{} with non-precessing spins. The search aims at finding an optimal waveform basis in the considered
parameter space (see \ac{sm}). 
For \ac{bns}, extrapolating our results, we find that 500 quasi-circular configurations 
may be sufficient to generate a waveform basis for this type of systems which is
accurate to $\bar{\cal{F}}\lesssim 0.3\%$. 
The greedy algorithm selects configurations 
with mass ratios up to $2$ and/or tidal deformabilities ranging from $\mathcal{O}(10)$ to $\mathcal{O}(10000)$.
Notably, \ac{nr} simulations with these parameters are entirely feasible and will be pursued in the future. 
For \ac{bhns}, a basis of $1000$ simulations may be enough to 
describe all the spin-aligned systems within the ranges considered up to $\bar{\cal{F}}\lesssim 3\%$.

\noindent {\textbf{\textit{Outlook}.}}
Being able to describe arbitrary orbits for generic compact binaries, 
\Dali{} poses itself as a milestone in the landscape of waveform models. 
\Dali{} is ready to use for GW inference with current observations, and can inform science cases for next generation
intereferometers, such as Einstein Telescope~\cite{Punturo:2010zz,Maggiore:2019uih}, 
Cosmic Explorer~\cite{Reitze:2019iox} and LISA~\cite{LISA:2017pwj}.
It can further support the interpretation of strong-field dynamics, as computed in \ac{nr} simulations, as for example in Refs.~\cite{Damour:2011fu,Hinderer:2013uwa,Damour:2014afa,Albanesi:2024xus} (see also \ac{sm} for an example).

Despite reaching a high level of accuracy, much work remains to improve \Dali{} across the binary parameter space. In the \ac{bbh} sector, ongoing work is focused on the development of \ac{nqc} for highly elliptic and hyperbolic mergers. A promising approach is the exploit the universality of the merger structure discovered in Ref.~\cite{Carullo:2023kvj}.
The ringdown model should also be generalized to arbitrary orbits, possibly 
including physical effects that are typically unmodeled, such as inspiral-inherited tail effects~\cite{Albanesi:2023bgi,DeAmicis:2024not,Islam:2024vro,DeAmicis:2024eoy,Ma:2024hzq} and
$m=0$ modes~\cite{Albanesi:2024fts}.
Absorption fluxes should also be improved in order to capture the \ac{bbh} phenomenology observed in simulations \cite{Chiaramello:2024unv}.
\Dali{} can currently generate full extreme-mass-ratio inspirals and mergers for non-precessing spins/planar orbits 
(not discussed here) using test-mass and \ac{gsf} information~\cite{Nagar:2022fep,Albertini:2022rfe,Albertini:2022dmc,Albertini:2023aol,Albertini:2024rrs,Albertini:2024agg}; 
extensions to arbitrary orbits are under development.
Further refinement of \ac{bns} and \ac{bhns} sector requires new sets of simulations as those indicated in the \ac{sm}. For example, a preliminary version of a \ac{nr}-informed tides is investigated in Ref.~\cite{Gamba:2023mww}.
Alternative routes to the approach presented here, are EOB models relying on post-Minkowskian
calculations~\cite{Damour:2022ybd,Rettegno:2023ghr,Buonanno:2024vkx,Buonanno:2024byg}, in particular the Lagrangian EOB approach 
recently proposed~\cite{Damour:2025uka}. 

\noindent {\textbf{\textit{Data availability.}}}
\Dali{} is publicly developed at {\footnotesize \url{https://bitbucket.org/teobresums/teobresums/src/Dali/}}.
A machine readable file with a list of simulations employed for the faithful analysis 
is distributed along this paper. Mismatches are computed using \texttt{PyART},
which is available at {\footnotesize \url{https://github.com/RoxGamba/PyART}}.

\noindent {\textbf{\textit{Acknowledgements.}}}
S.A. and R.G. contributed equally to this work.
S.A. and S.B. acknowledge support from the Deutsche Forschungsgemeinschaft (DFG) project ``GROOVHY'' (BE 6301/5-1 Projektnummer: 523180871).
R.G. acknowledges support from NSF Grant PHY-2020275 (Network for Neutrinos, Nuclear Astrophysics, and Symmetries (N3AS)).
S.B. and J.F. acknowledge support by the EU Horizon under ERC Consolidator Grant, no. InspiReM-101043372.
A.G. acknowledges support by the Deutsche Forschungsgemeinschaft (DFG) under Grant No. 406116891 within the Research Training Group RTG 2522/1 and the 
Comunitat Autònoma de les Illes Balears through the Conselleria d'Educació i Universitats with funds from the European Union - NextGenerationEU/PRTR-C17.I1 (SINCO2022/6719).
The authors sincerely thank
D.~Chiaramello, M.~Panzeri, and P.~Rettegno for the recent developments of \Dali{}, and
T.~Andrade and J.~Trenado for providing access to the \texttt{ICC} catalog data.
S.A. thanks A.~Buonanno and L.~Pompili for suggestions on the draft.  
The authors also express their gratitude to Gaynor Sullivan for her support during the final stages of this work.
S.A. is grateful to Merlino Capobianco for useful discussions. 
R.G. acknowledges A.~Overton and A.~Scarrone for inspiration and support.

Simulations were performed on SuperMUC-NG at the Leibniz-Rechenzentrum (LRZ) Munich and and on the national HPE Apollo Hawk at the High Performance Computing Center Stuttgart (HLRS).
The authors acknowledge the Gauss Centre for Supercomputing e.V. (\url{www.gauss-centre.eu}) for funding this project by providing computing time on the GCS Supercomputer SuperMUC-NG at LRZ (allocations {\tt pn36go}, {\tt pn36jo} and {\tt pn68wi}). The authors acknowledge HLRS for funding this project by providing access to the supercomputer HPE Apollo Hawk under the grant number INTRHYGUE/44215 and MAGNETIST/44288.
Computations were also performed on the ARA cluster at Friedrich Schiller University Jena and on the {\tt Tullio} INFN cluster at INFN Turin. The ARA cluster is funded in part by DFG grants INST 275/334-1 FUGG and INST 275/363-1 FUGG, and ERC Starting Grant, grant agreement no. BinGraSp-714626.

\bibliography{refs20250318.bib,refs_loc20250318.bib}

\section{Supplemental Material}

\begin{figure}[t]
  \centering 
  \includegraphics[width=0.48\textwidth]{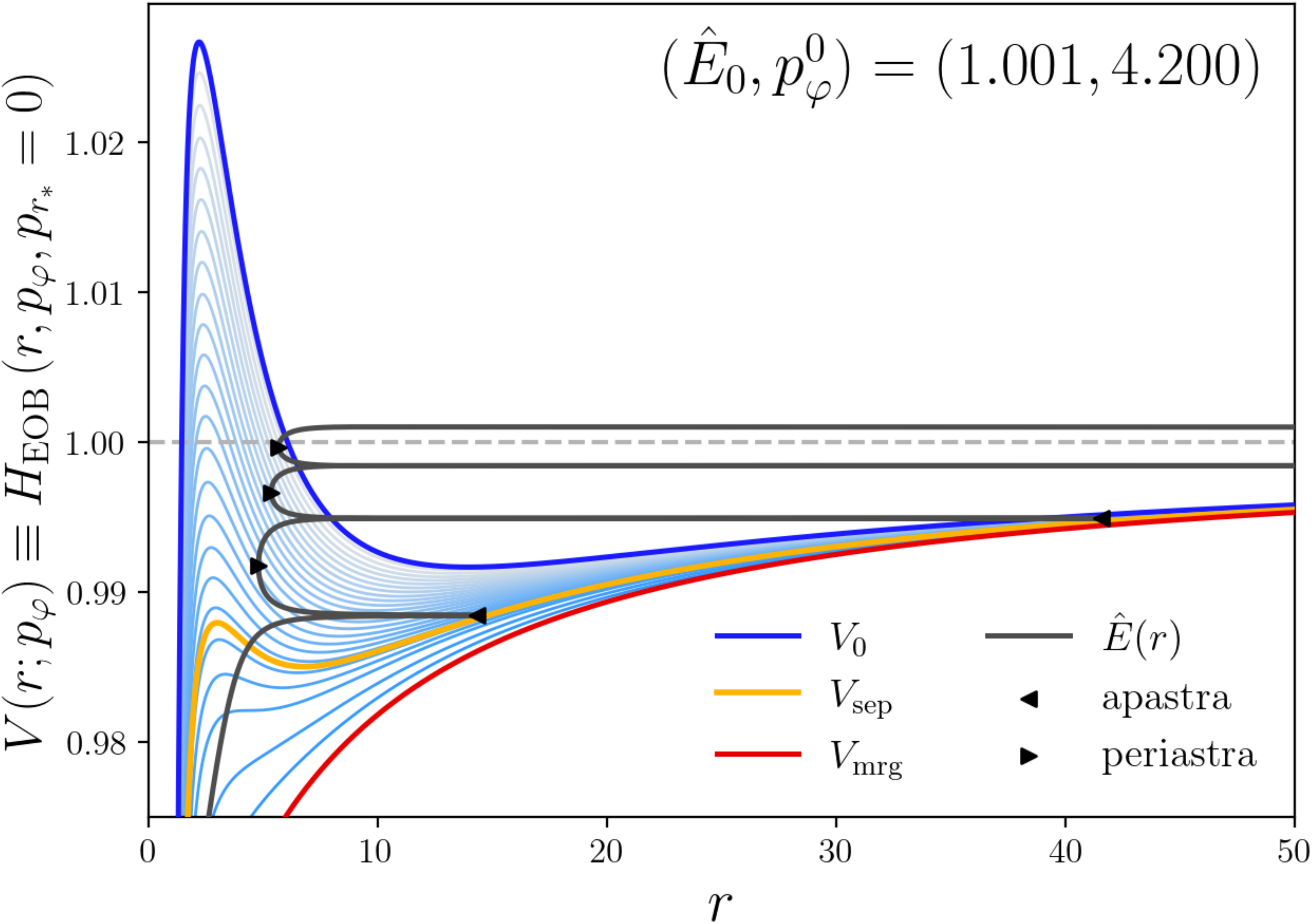}
  \includegraphics[width=0.48\textwidth]{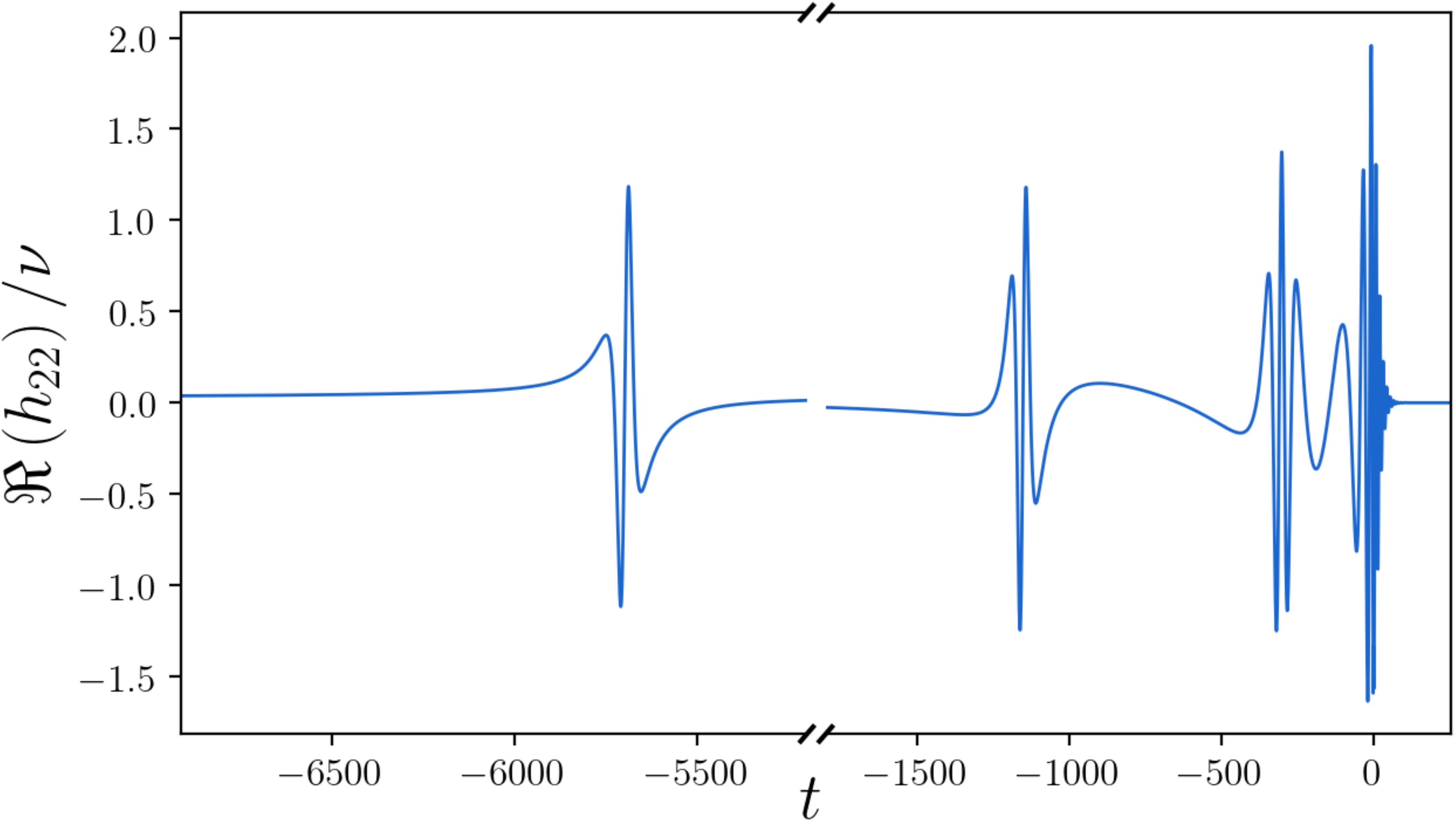}
  \caption{Upper panel: EOB potential for an equal-mass non-spinning binary. Lower panel: corresponding (2,2) waveform. 
  The peaks before merger correspond to the periastra passages.}
  \label{fig:eob_potential}
\end{figure}

\noindent {\textbf{\textit{\ac{eob} analysis of a dynamical capture.}}}
We illustrate how the dynamics of a general relativistic two-body system can be interpreted in terms
of the EOB radial potential. This effective potential can be
defined from the \ac{eob} Hamiltonian by setting the radial momentum to zero, $V(r;p_\varphi) \equiv H_{\rm EOB} (r,p_\varphi;p_{r_*}=0)$. 
Note that this potential evolves over time due to the loss of angular momentum of the binary, $p_\varphi$. In Fig.~\ref{fig:eob_potential}
we report this evolution for a dynamical capture, together with the radial evolution of the energy, $E(r)$ (black line). 
During the first close encounter, enough radiation is emitted
to make the system bound ($E<M$). 
As a consequence, after the first close encounter the two radial turning points, the 
apastron $r_+$ and the periastron $r_-$, are naturally identified by the condition $E=V(r_\pm)$. 
However, at separatrix-crossing time, the local maximum of the potential, $V_{\rm max}$,
becomes smaller than the energy of the binary. From this instant\footnote{Note that this is 
a generalization for generic mass ratio of the concept of 
separatrix  for orbits in Kerr spacetime~\cite{OShaughnessy:2002tbu,Stein:2019buj}.}, the periastron is no longer defined and the system will plunge
at the next close encounter. In Fig.~\ref{fig:eob_potential} we also report the corresponding (2,2) waveform,
which exhibits bursts of radiation in correspondence of the periastron passages and the merger. 
This allows direct connection between the dynamics and the waveform. 

Moreover, the definition of the potential also allows a definition of an eccentricity and a semilatus rectum in terms of 
radial turning points, $e_{\rm EOB} = (r_+-r_-)/(r_++r_-)$ and $p_{\rm EOB} = 2 r_+ r_-/(r_++r_-)$, 
which can be useful to gain intuition of the dynamics considered. However, since these quantities are 
gauge-dependent and not defined after the separatrix-crossing
time, they are not a suitable choice to characterize the late-evolution of the dynamics, especially during
the plunge and the merger. Indeed, for the late-dynamics is convenient to use quantities defined in terms 
of energy and angular momentum, which are instead defined through the whole evolution of the binary\footnote{In other words, 
the map between the (conservative) constants of motion $(e_{\rm EOB},p_{\rm EOB})\leftrightarrow (E, p_\varphi)$ is valid
until the separatrix-crossing time, after which  $(e_{\rm EOB},p_{\rm EOB})$ are no longer defined, as opposed to $(E, p_\varphi)$.},
as extensively discussed in Ref.~\cite{Albanesi:2023bgi,Carullo:2023kvj}.

\noindent {\textbf{\textit{\CoRe{} \ac{bhns} dataset.}}}
Our faithfulness analysis also considered a new set of \CoRe{} simulations performed
using \texttt{BAM}~\cite{Brugmann:2008zz,Thierfelder:2011yi}.
This dataset contains 13 quasi-circular spin-aligned
\ac{bhns} configurations and one precessing binary, with mass-ratios ranging from 1.9 to 4.1.
Among them, there are two high resolution runs with more than 10 orbits,
including the precessing configuration. These data will be discussed in 
Ref.~\cite{Gonzalezinprep} and released on the \CoRe{} database~\cite{Gonzalez:2022prs}.

\noindent {\textbf{\textit{Scatterings.}}}
% Scatterings
%
\begin{figure}[t]
  \centering
    \includegraphics[width=0.48\textwidth]{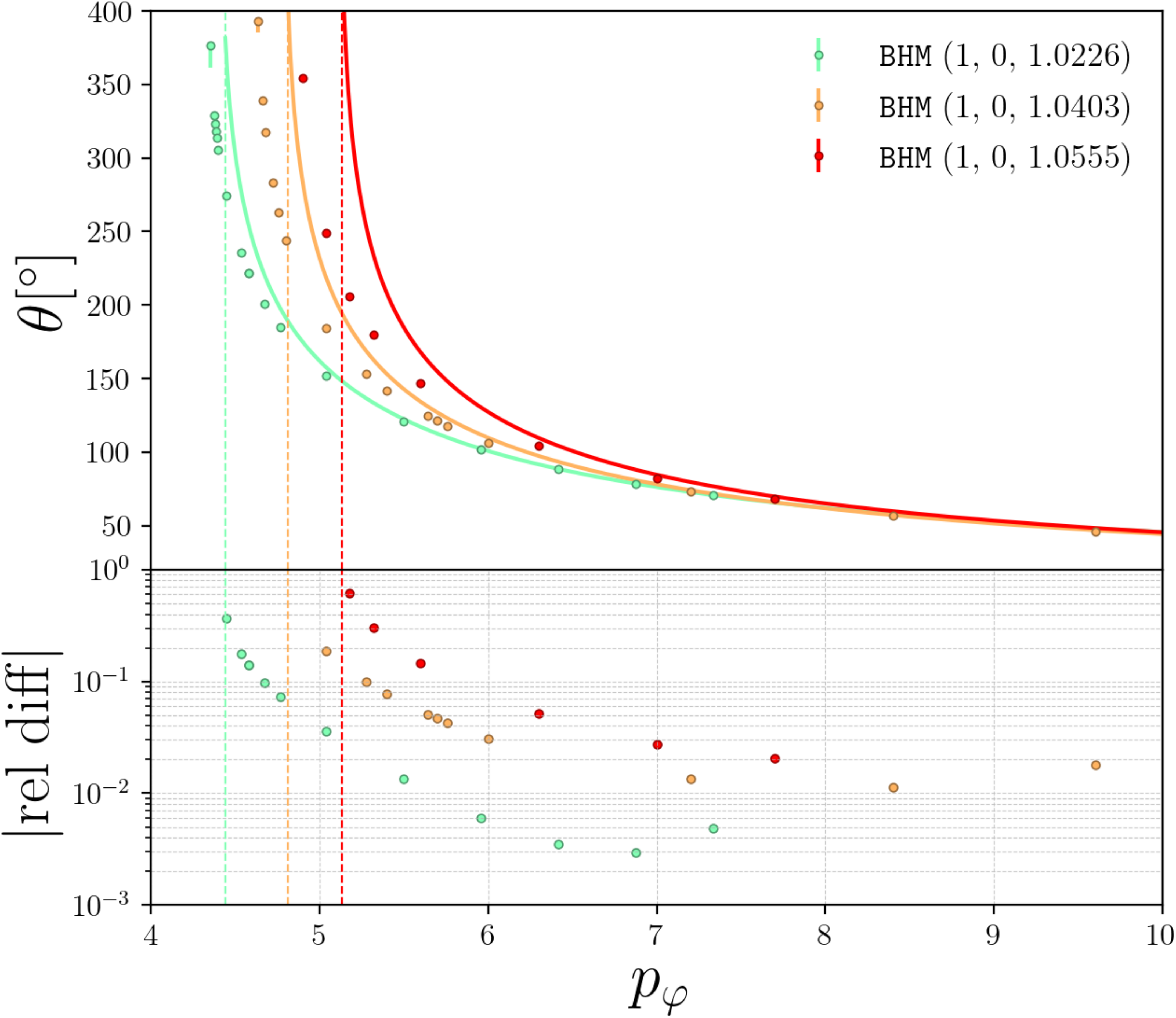}
    \includegraphics[width=0.48\textwidth]{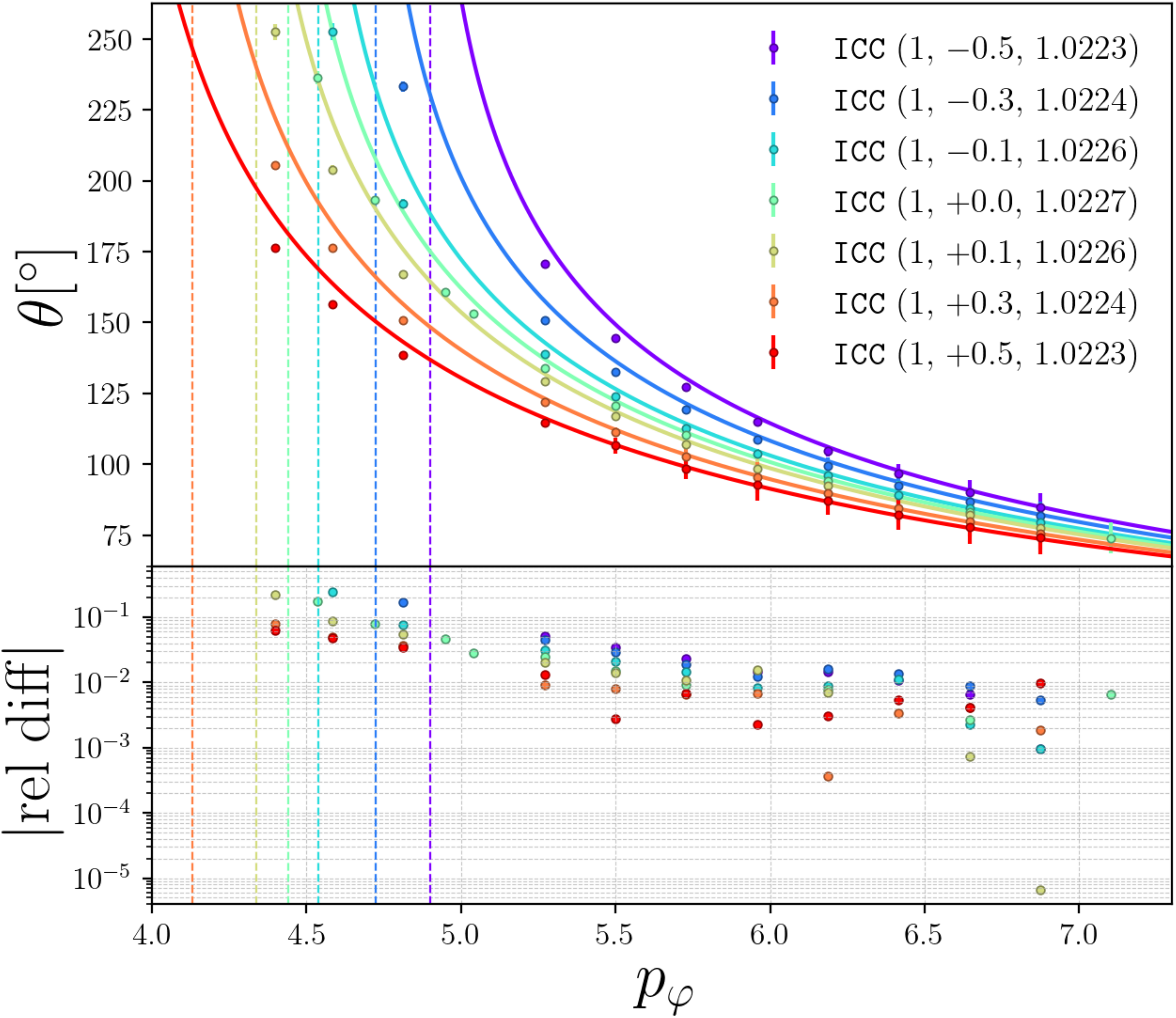}\\
    \caption{Equal-mass scattering configurations from Ref.~\cite{Rettegno:2023ghr} (upper plot) 
    and Ref.~\cite{Fontbute:2024amb} (lower plot).
    The predictions by \Dali{} are reported with solid lines. The vertical dashed lines correspond to the predicted 
    transition from scattering to capture. We report the EOB/NR relative differences in the subpanels.
    The colors highlight different initial energies or different spin configuration, $(q,\chi_{1,2}, E_0/M)$, as specified in the legends.
    }
  \label{fig:scat}
\end{figure}
For unbound systems, another gauge invariant quantity that can be considered to assess the accuracy of the EOB 
approach~\cite{Damour:2014afa,Hopper:2022rwo,Albanesi:2024xus,Buonanno:2024vkx}, as well as \ac{pn} and post-Minkowskian 
calculations~\cite{Damour:2022ybd,Rettegno:2023ghr}, is the scattering angle.
We thus proceed to consider \ac{nr} results from recent works~\cite{Rettegno:2023ghr,Fontbute:2024amb} to evaluate the 
accuracy of \Dali{}. EOB/NR comparisons for different initial energies in the equal-mass non-spinning case~\cite{Rettegno:2023ghr} are reported in the 
upper panel of Fig.~\ref{fig:scat}, while equal-mass configurations at fixed initial energy $E_0\simeq 1.022\,M$
and spins in the range $\chi_1=\chi_2 \in [-0.5, 0.5]$ are reported below\footnote{Note that here we do not perform any optimization 
on the initial data used to start the EOB evolution, as opposed to the mismatch calculation.}~\cite{Fontbute:2024amb}. 
We also show, with vertical dashed lines, the transition from scattering to capture as 
predicted by \Dali{}. While the model is accurate for low scattering angles or, equivalently, for large initial angular momenta,
the accuracy decreases for large scattering angles, \ie~close to the transition from scattering to capture. Notably,
previous versions of the model~\cite{Nagar:2024dzj} showed smaller relative difference close to the transition region, highlighting that 
different resummation schemes in the radiation reaction can have a significant impact on the strong-field dynamics. 

\noindent {\textbf{\textit{Greedy algorithm.}}}
Numerical simulations of \ac{bns} and \ac{bhns}
mergers are comparably more limited than \ac{bbh} in terms of
parameter space coverage and number of pre-merger orbits
simulated. These simulations require at very least
general-relativistic hydrodynamics and a choice of
\ac{eos}. Therefore, they are computationally more challenging and the  
coverage of the parameter space (masses, spins, tidal polarizability
parameters) requires a modeling choice for the matter
composition. Further, waveform's uncertainties increase significantly with the
number of simulated orbits, see \eg~\cite{Bernuzzi:2011aq,Bernuzzi:2012ci}.
Current waveform models for these system are designed using a limited,
${\cal O}(50)$, number of simulations. 

We use \Dali{} and an iterative greedy algorithm to identify the most
informative \ac{bns} and \ac{bhns} configurations to simulate in the
near future. The Greedy iterate~\cite{Field:2011mf,Canizares:2013ywa}
augments the basis with the next orthogonal element by picking a
furthest point in the parameter space from the basis.
Our algorithm is similar to the procedure employed to build reduced
order models~\cite{Field:2011mf,Canizares:2013ywa}.
The main difference is that, instead of an overlap-based metric, we
employ a mismatch-based one to identify points in the parameter
space. This allows us to immediately maximize over time and phase
shifts, which are irrelevant for modeling purposes.

We explore the \ac{bns} parameters space in the following ranges:
$m_{1,2} \in [1,2] M_{\odot}$, $\chi_{1,2} \in [-0.2, 0.2]$, and
$\Lambda_{1,2}$ determined by masses and a sample of 10
piecewise polytropic \ac{eos}: 
2B, 2H, ALF2, APR4, ENG, H4, MPA1, MS1, MS1b, SLy~\cite{Read:2008iy}.
The algorithm is initialized with $25$ of the high-resolution \texttt{BAM}
simulations used in the main text, and proceeds as follows: 
\begin{enumerate}
\item For each existing simulation, we compute the \ac{eob}
  waveform and store it in the ``known basis'' set, $\mathcal{B}$. 
\item We then randomly sample $N = 1000$ points in the parameter
  space, and for each point compute the \ac{eob} waveform $h_i$
  and its mismatch with respect to each element $h_j$ in
  $\mathcal{B}$, $\mathcal{\bar{F}}^i_j$. The residual $r_i$ between
  $h_i$ and the basis is defined as the minimum mismatch over all
  elements in $\mathcal{B}$: 
  \begin{equation}
  r_i = \min_j \mathcal{\bar{F}}^i_j
  \end{equation}
\item We select the element with the largest residual over the set
  of $N$ points, and add it to $\mathcal{B}$. 
\item We repeat steps 2-3 until we have reached a desired number
  of simulations (in our case, $N_{\text{sim}} = 100$). 
\end{enumerate}
EOB waveforms are computed from a (dimensionless) starting frequency
of $f_0 = 0.0049$ ($0.0055$) for \acp{bns} (\acp{bhns}), and the mismatch computed with flat \ac{psd} within the
frequency range $[0.0055, 0.04]$ ([$0.007, 0.1]$). 
This \ac{bns} range roughly corresponds to systems spanning about $3000 M$ before merger, 
mimicking the frequency range typically covered by \ac{nr} simulations.

\begin{figure*}[t]
  \centering 
  \includegraphics[width=0.48\textwidth]{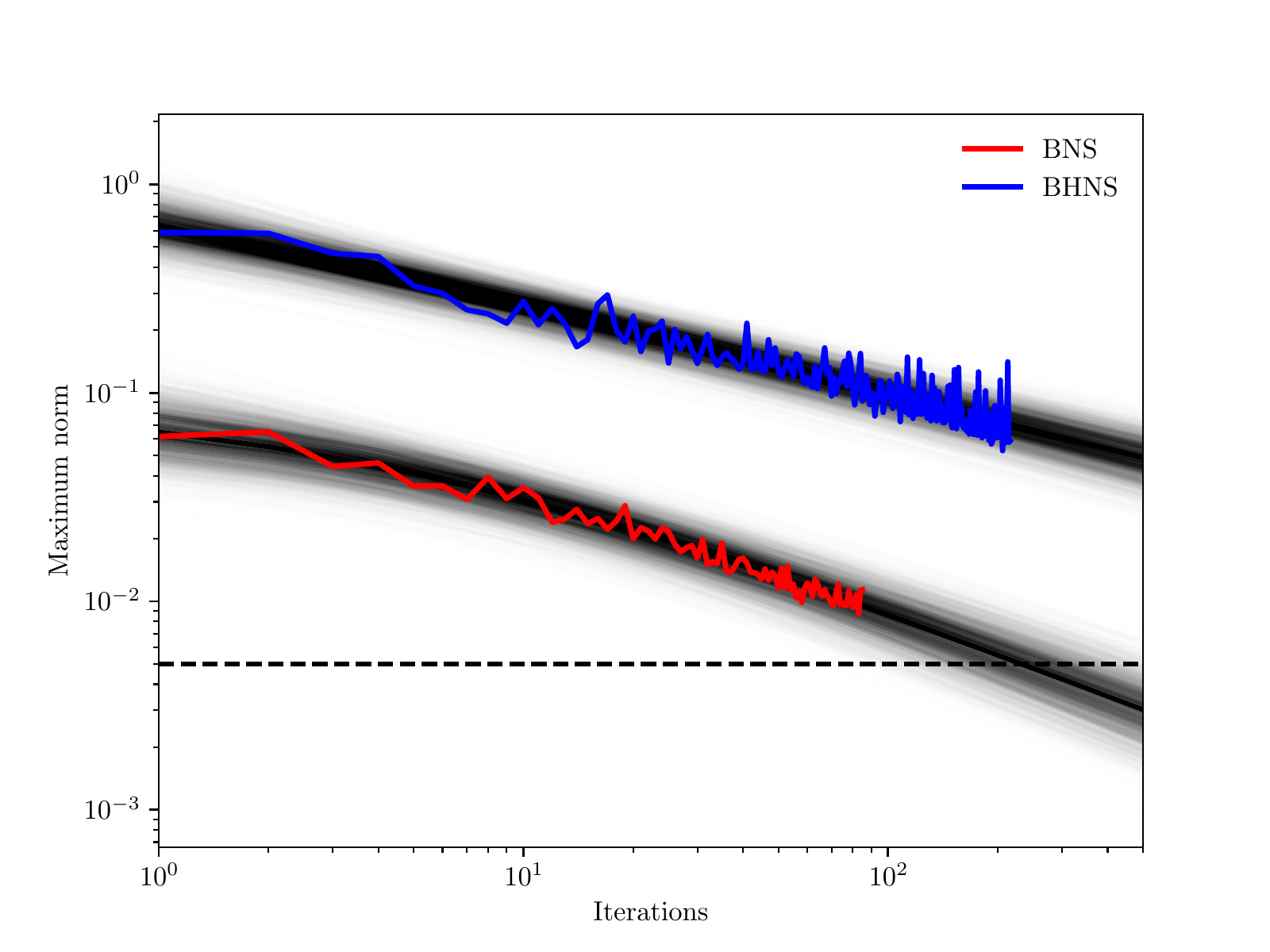}
  \includegraphics[width=0.48\textwidth]{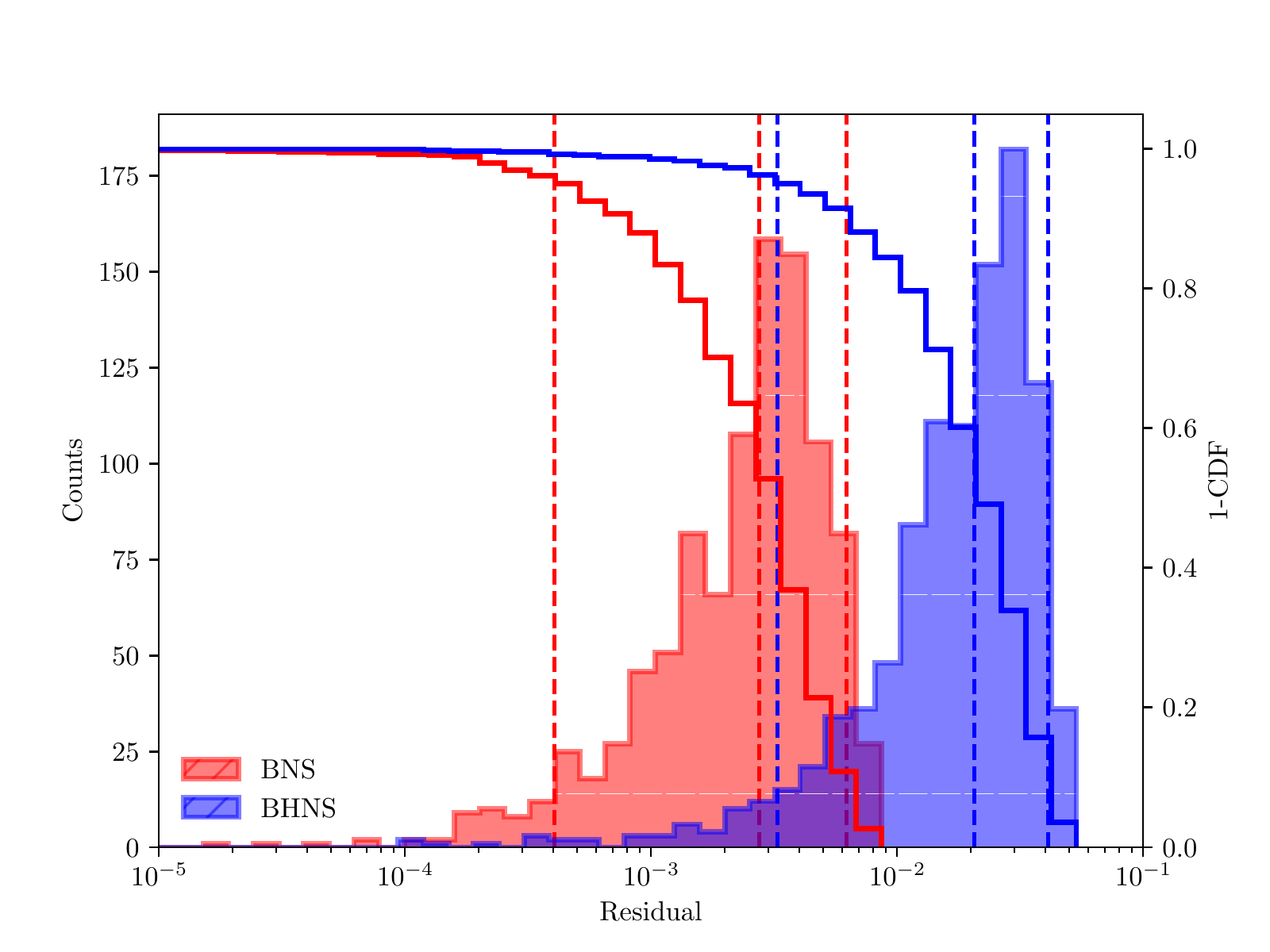}
  \caption{Greedy search for \ac{bns} and \ac{bhns}. Left: Evolution of the residual as a function of the iteration. 
    For \acp{bns}, the residual halves from $\approx 3\%$ to $\approx 1.5\%$ in the first $50$ iterations, reaching $\sim 0.8\%$ after $100$
    steps. This indicates the success of the greedy algorithm in identifying the most informative points in the parameter space.
    For \acp{bhns}, the parameter space is wider and residual larger and more slowly decreasing, reaching $\approx 7\%$ after $200$
    iterations. Right: Distribution of residuals of $1000$ random points sampled within the \ac{bns} (red) and \ac{bhns} (blue) parameter space after the completion of the greedy algorithm.
    The distribution of \ac{bns} (\ac{bhns}) residuals has a tail extending up to mismatches of $\sim 8 \times 10^{-3}$ ($\sim 0.05$), 
    while $50\%$ of distribution lies below $0.003$ ($0.02$).}
  \label{fig:greedy_dist}
\end{figure*}

Figure~\ref{fig:greedy_dist} (left panel) shows the convergence of the
residual as iteration proceed. 
Note we select the number of simulations as stopping criterion, but
alternatively a minimum mismatch could also be chosen.
After the completion of the greedy algorithm we selected an additional
$1000$ random points in the parameter space and computed their
residual from the basis. These are shown in the right panel of Figure
\ref{fig:greedy_dist}. 
The distribution of residuals has a tail extending up to mismatches of
$\sim 8 \times 10^{-3}$, while the bulk of the distribution lies well
below this value ($\sim 2-4 \times 10^{-3}$). This suggests that the
greedy algorithm is able to identify the most informative points in
the parameter space, as desired.

\begin{figure*}[t]
  \centering 
  \includegraphics[width=0.48\textwidth]{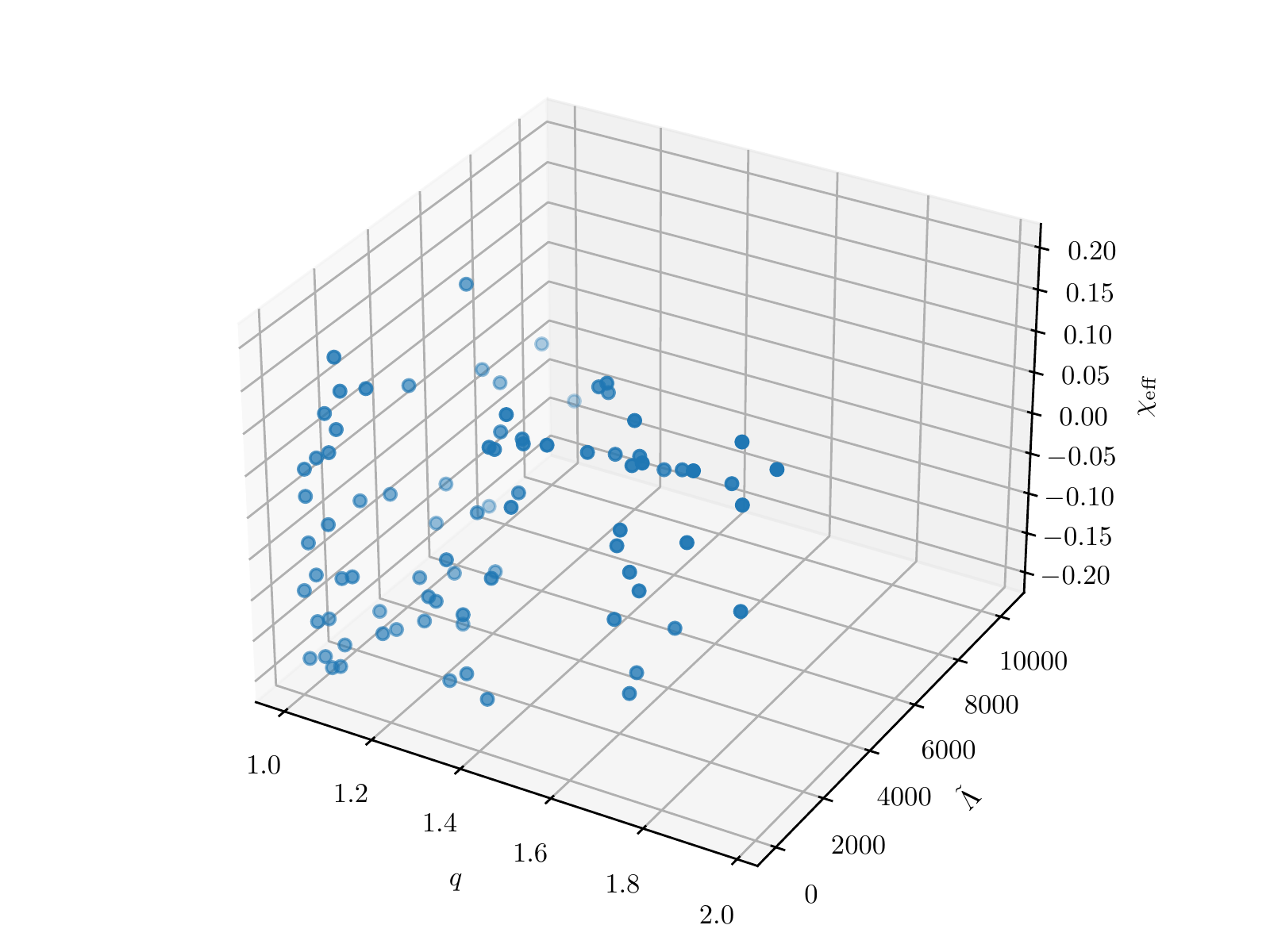}
  \includegraphics[width=0.48\textwidth]{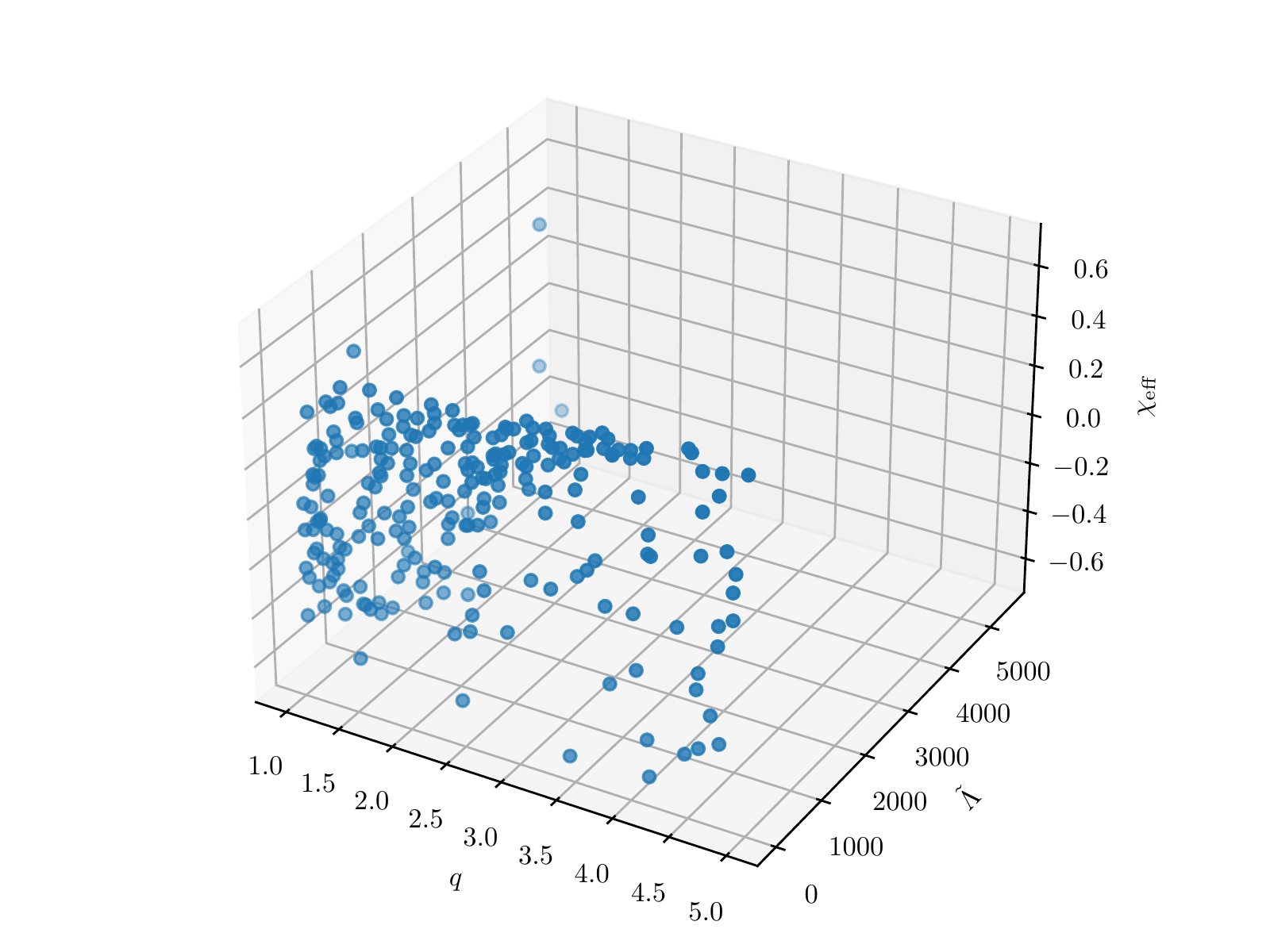}
  \caption{Distribution of greedy basis waveforms for \ac{bns} (left) and \ac{bhns} (right) 
  visualized in the space of mass-ratio, effective spins and reduced tidal parameter.}
  \label{fig:greedy_basis}
\end{figure*}

A similar procedure is applied for \ac{bhns}. The final results are
shown in Fig.~\ref{fig:greedy_basis}, where the waveform basis are
visualized in the space of mass-ratio, effective spins and reduced
tidal parameter. Performing these simulations is computationally
expensive but it is feasible with current \ac{nr} codes. Future work
will focus on performing these simulations at multiple resolutions and
adopting them as basis for systematic comparisons and improvements of
\Dali{}. 

\end{document}